\documentclass[12pt,a4paper]{article}

\usepackage{latexsym,graphicx}

\title{A New Look to Massive Neutron Cores }

\author{Ll.\ Bel\\
\emph{Fisika Teorikoa, Euskal Herriko Unibertsitatea}, \\
\emph{P.K. 644, 48080 Bilbo, Spain}
}

\begin{document}
\maketitle

\begin{abstract}

We reconsider the problem of modelling static spherically symmetric
perfect fluid configurations with an equation of state from a point of
view of that requires the use of the concept of principal transform of a
3-dimensional Riemannian metric. We discuss from this new point of view
the meaning of those familiar quantities that we call density, pressure
and geometry in a relativistic context. This is not simple semantics. To
prove it we apply the new ideas to recalculate the maximum mass that a
massive neutron core can have. This limit is found to be of the order of
3.8\,$M_\odot$ substantially larger than the Oppenheimer and Volkoff
limit.

\end{abstract}

\section*{Introduction}

We review in Section 1 the basic equations of the models being
considered as well as the concept of principal transform of a
3-dimensional Riemannian metric which is at the core of our new point of view 
to understand these models. 

Section 2 is devoted to lay down the fundamental system of equations to
be integrated. We use spherical space coordinates of the quo-harmonic
class which allow to implement ${\cal C}^1$ class smoothness across a
sharp border when there is one.

In Section 3 we define the concept of proper mass $M_p$ to be used to
define the binding energy $E_b$ of the models as the difference between
the active gravitational mass $M_a$ and $M_p$. We define also the
concept of proper mass density which is a fundamental hybrid concept
related to $M_p$ and the principal transform of the quotient space metric. Its
relation with the pressure $p$ characterizes the fluid source
independently of the solution being considered.

Section 4 is devoted to establish that the binding energy $E_b$ can be
obtained as the integral over all space of a localized energy density
$\sigma$ which depends only on some of the gravitational potentials 
and its first radial derivatives.

In Section 5 we linearize the fundamental system of equations to obtain
the linearized expression of the binding energy $E_b$, which coincides
with the familiar Newtonian one, thus providing a partial justification to the
definition of $E_b$ in the non linear regime. We obtain also the
linearized expression of the energy density $\sigma$.

The last section contains our proposed application of the new point of
view to the study of massive neutron cores. The equation of state is the
usual one for a degenerate neutron gas, but both the density and the
pressure are variants of those used by Oppenheimer and Volkoff
\cite{OppenVolkoff}. Our main result is that the maximum limit mass that
a neutron core can have is approximately 3.8\,$M_\odot$ 
instead of 0.7\,$M_\odot$ in \cite{OppenVolkoff}. 

\section{Static Spherically Symmetric Models}

We shall be interested in this paper on global spherically symmetric
models, which we shall write using a time adapted coordinate
and polar-like space coordinates:

\begin{equation}
\label {1.1}
ds^2=-A^2(r)c^2dt^2+d\hat s^2
\end{equation}
where:

\begin{equation}
\label {1.2}
d\hat s^2=B^2(r)dr^2+B(r)C(r)r^2d\Omega^2, \quad
d\Omega^2=d\theta^2+\sin^2\theta d\varphi^2, 
\end{equation}
solution of Einstein's field equations:

\begin{equation}
\label {1.3}
S_{\alpha\beta}=-\chi T_{\alpha\beta},\quad \chi=8\pi G/c^4
\end{equation} 
where the r-h-s describes a compact fluid source, or with fast decreasing density, 
with two ``flavours": isotropic or a special kind of anisotropic
pressure to be presented in a moment.

The quotient 3-dimensional metric (\ref{1.2}) can be written using a variety of
supplementary coordinate conditions belonging to two different types:
algebraic or differential. The most often used is the curvature
condition which uses a radial $\tilde r$ coordinate with the algebraic
condition:

\begin{equation}
\label {1.4}
\tilde C=\tilde B^{-1}
\end{equation}   
We shall refer also to any other quantity which assumes the use of this radial
coordinate with a tilde overhead.

We shall use here almost exclusively the quo-harmonic condition which 
restricts the $r$ coordinate with the differential condition:

\begin{equation}
\label {1.5}
C^\prime=\frac{2}{r}(B-C)
\end{equation}  
where the prime means derivative with respect to r. 
The use of a differential condition makes possible the
construction of global ${\cal C}^1$ models with a sharp boundary with
vacuum, something which is not possible when using the coordinate $\tilde r$.

Notice that if $B(r)$ and $C(r)$ are known the curvature coordinate
$\tilde r$ is simply the following function of $r$:

\begin{equation}
\label {1.9}
\tilde r=\sqrt{B(r)C(r)}r
\end{equation}
while, on the contrary, to obtain explicitly the inverse function is
in most cases of interest impossible or very cumbersome.
 
Despite the emphasis we put on the use of the quo-harmonic coordinate
$r$ let us be clear from the beginning that the main conclusion of
this paper will not owe anything to a particular choice of
coordinates. It will owe instead all to a new concept: that of a
principal transformation of a 3-dimensional Riemannian metric such
as (\ref{1.2}).

By definition, in the particular case we are considering, the
principal transform of (\ref{1.2}) is a new 3-dimensional Riemannian 
metric\,\footnote{See for instance references \cite{Bel96} or \cite{AguiBel01}}:

\begin{equation}
\label {1.6}
d\bar s^2=\Phi^2(r)B^2(r)dr^2+\Psi^2(r)B(r)C(r)r^2d\Omega^2
\end{equation}    
such that: i)

\begin{equation}
\label {1.7}
\bar R^i_{jkl}=0, \quad i,j,k,l=1,2,3
\end{equation}
and: ii)

\begin{equation}
\label {1.8}
(\hat\Gamma^i_{jk}-\bar\Gamma^i_{jk})\hat g^{jk}=0, 
\end{equation}
where, with otherwise obvious notations, the quantities with a hat
overhead refer to the metric (\ref{1.2}) and the quantities with a bar
overhead refer to the metric (\ref{1.6}). Notice that both conditions
above being tensor conditions under any transformation of space
coordinates the concept of Principal transformation is intrinsic to
the Killing time congruence we are considering. 

One of the practical conveniences of using polar quo-harmonic space
coordinates from the outset in the very process of model-building is
that, when appropriate boundary conditions are taken into account,
the principal transform of (\ref{1.2}) is just:

\begin{equation}
\label {1.10}
d\bar s^2=dr^2+r^2d\Omega^2
\end{equation}   
and:

\begin{equation}
\label {1.11}
\Phi(r)=1/B(r), \quad \Psi(r)=1/\sqrt{B(r)C(r)}
\end{equation}

We shall consider two types of energy-momentum tensors: i) those
describing standard perfect fluids with isotropic pressure:

\begin{equation}
\label {1.12}
T_{00}=\rho A^2, \quad T_{ij}=p\hat g_{ij}, \quad (x^0=ct)
\end{equation}
and a new type of fluid where the isotropy of the pressure is meant
in the sense of the principal transform (\ref{1.6}):

\begin{equation}
\label {1.13}
T_{00}=\rho A^2, \quad T_{ij}=p\bar g_{ij}
\end{equation} 
i.e. in the sense of (\ref{1.10}) if polar quo-harmonic coordinates are
used.

Notice that this second case, that we shall present in Section 3 as
being the truly isotropic fluid, can be considered from the standard
point of view as an anisotropic fluid of a particular type. Namely
one for which the radial and tangential pressures are related to a
single function $p(r)$ as follows:

\begin{equation}
\label {1.14}
p_r=pB^{-2}, \quad p_t=p(BC)^{-1}
\end{equation} 
We can deal with both cases at once introducing in the field
equations a two-valued flag $f=0$ or $f=1$ and the following
expressions for $p_r$ and $p_t$:

\begin{equation}
\label {1.15}
p_r=(fB^{-2}+1-f)p, \quad  p_t=(f(BC)^{-1}+1-f)p
\end{equation}
thus $f=0$ corresponds to the standard case and $f=1$ corresponds to the
new case.

As usual we shall make use of a compressibility equation:

\begin{equation}
\label {1.16}
\rho=\rho(p) \quad \hbox{or} \quad p=p(\rho) 
\end{equation}
depending on convenience, to describe the physics of the source.

The boundary conditions at infinity will be:

\begin{equation}
\label {1.17}
\lim_{r\to\infty}\Xi=1, \quad  \lim_{r\to\infty}r^3\Xi^\prime=0 
\end{equation}
where $\Xi=A, B$ or $C$; and the regularity conditions at the centre of symmetry of the 
configuration will be:

\begin{equation}
\label {1.18}
\Xi^\prime_0=0,  \quad \Xi^{\prime\prime}_0<\infty
\end{equation}

Finally, in those cases where the model has a sharp boundary between
an interior with $\rho> 0$ and vacuum we shall require the
continuity of both $Xi$ and $X^\prime$, so that the space-time metric
will be of global ${\cal C}^1$ class. 

The system of units we shall use throughout will be such that:

\begin{equation}
\label {1.19}
\bar c=1, \quad \bar G=1
\end{equation}
where $\bar c$ is the speed of light in vacuum and $\bar G$ is Newton's
constant. A more specific system of units of this class will be chosen in Sect. 6.

\section{Explicit field equations}

Taking into account the coordinate condition (\ref{1.5}) and also its
derivative:

\begin{equation}
\label {2.3}
C^{\prime\prime}=2\frac{B^\prime}{r}-\frac{6}{r^2}(B-C)
\end{equation}
the field equation $S_{00}=\rho A^2$ can be written as follows:

\begin{equation}
\label {2.1}
B^{\prime\prime}=-3\frac{BB^\prime}{rC}-\frac{B^2}{r^2C}+\frac54\frac{
{B^\prime}^2}{B}+\frac{B^3}{r^2C^2}-8\pi\rho B^3
\end{equation} 
The field equation $S_{22}=-p_tBCr^2$ (or $S_{33}=-p_tBCr^2\sin^2\theta$) can be
written using the preceding Eq. (\ref{2.1}) and the second Eq.
(\ref{1.15}) as:

\begin{eqnarray}
\label {2.2}
A^{\prime\prime}=&& \frac{1}{rC}\left(\frac12 AB^\prime
-BA^\prime\right)+\frac12\frac{BA}{r^2C}\left(\frac{B}{C}-1\right) \nonumber
+\frac18\frac{A{B^\prime}^2}{B^2}+\frac12\frac{A^\prime B^\prime}{B}\\
&&+4\pi\rho B^2A+8\pi pAB\left(f/C+B(1-f)\right)
\end{eqnarray}

The remaining equation we have to take care of is $S_{11}=-p_rB^2$, 
or taking into account (\ref{1.5}):

\begin{equation}
\label {2.2bis}
\frac{B}{r^2C}-\frac14\frac{{B^\prime}^2}{B^2}-\frac{B^2}{r^2C^2}-
\frac{A^\prime B^\prime}{BA}-\frac{B^\prime}{rC}-2\frac{BA^\prime}{rCA}=
-8\pi p(f+(1-f)B^2)
\end{equation}

On the other hand from the conservation equation:

\begin{equation}
\label {2.4}
\nabla_\alpha S^\alpha_1=0 \quad (x^1=r)
\end{equation}
or:

\begin{equation}
\label {2.6}
S^{\prime 1}_1+\Gamma^0_{10}(S^1_1-S^0_0)+\Gamma^2_{12}(S^1_1-S^2_2)+
\Gamma^3_{13}(S^1_1-S^3_3)=0
\end{equation}
and Eqs. (\ref{1.15}) we derive the equation:

\begin{equation}
\label {2.5}
p^\prime=-\frac{A^\prime}{A}\left[\rho(fB^2+1-f)+p\right]+f\left[2\frac{B^\prime}{B}-
\left(\frac{B^\prime}{B}+\frac{2}{r}\frac{B}{C}\right)\left(1-\frac{B}{C}\right)\right]p
\end{equation}

Taking into account the regularity conditions (\ref{1.18}) at the
centre of symmetry of the configuration in the preceding equations 
it is easy to see that they imply:

\begin{equation}
\label {2.7}
B_0=C_0, \quad p_{t0}= p_{r0},  \quad S_{110}=-B_0^2p_{t0}
\end{equation} 
Therefore the remaining field equation $S_{11}=-p_rB^2$ is satisfied at
the origin $r=0$ and from from (\ref{2.1}),
(\ref{2.2}) and (\ref{2.6}) it follows that it is satisfied
everywhere.

What follows is the summary of this section and the preceding one: The models we are
considering will be fully described by the field variables 
$A, B, C$ and the source variables $\rho, p$; the latter being
related by a compressibility equation of either type (\ref{1.16}).
This
complete set of variables is constrained to satisfy the system of
differential equations (\ref{1.5}), (\ref{2.1}), (\ref{2.2}) and
(\ref{2.5}).
Appropriate initial conditions will be $A_0$, $B_0=C_0$ and
$\rho_0>0$ (or $p_0>0$).The initial values of $A_0$ and $B_0$ have to
be chosen such that the asymptotic conditions (\ref{1.17}) are
satisfied. The boundary of the source will be defined by the first
zero $r=R$ of the pressure $p$, and beyond the vacuum field
equations will be required. The continuity of $A$, $B$ and $C$ and
its first derivatives is automatically implemented across the boundary of the source.

\section{Physical and geometrical interpretations}

This will be the more difficult section of this paper, although it does
not contain any calculation, because it deals about the meaning of
words of common use. 

When we look at Eqs. (\ref{1.3}) as equations to be solved we all refer
to the r-h-s as the source term. But this is not quite correct
because the energy-momentum tensor depends on the coefficients of the
unknown metric. In the case we are considering in this paper the real
source variables are the so called density $\rho$ and pressure $p$ related
by a compressibility equation (\ref{1.16}). The meaning of these three
ingredients deserve to be examined with some detail. A density by
definition is a mass per unit volume, and a pressure is a force per
unit surface. Therefore to be clear about them we must tell of what
mass are we talking about and to what geometry of space are we
referring when using the words volume and surface. 

As we all know the concept of mass is tricky because it comes in
three flavours:  inertial mass, passive gravitational mass and active
gravitational mass. Newtonian theory assumes the proportionality of
the three masses and General relativity assumes the proportionality of inertial and
passive gravitational masses of test bodies. Beyond that we have a
few decisions to be taken.

To decide what geometry of space to use  is also a tricky problem in
relativity theory because we have to decide whether this geometry has
to be known before we solve the field equations or will be known only
after they have been solved, in which case the meaning of $\rho$, 
$p$ and the compressibility Eq. (\ref{1.16}) will also be known only after 
the problem has been solved.

Of the three types of mass, only the meaning of active gravitational
mass was settled very early by Tolman \cite{Tolman} identifying it
with the Newtonian mass at infinity and proving that it can be
calculated as the following integral over the source:

\begin{equation}
\label {3.1}
M_a=4\pi\int_0^\infty{\rho{\tilde r}^2}\, d\tilde r
\end{equation}     
where we remind that $\tilde r$ is the curvature radial coordinate. Using
(\ref{1.9})
this formula can be written equivalently as:

\begin{equation}
\label {3.2}
M_a=4\pi\int_0^\infty{\rho F(r)r^2}\, dr
\end{equation} 
where:

\begin{equation}
\label {3.2.1}
F(r)=\frac12 \sqrt{BC}(B^\prime Cr+2B^2)
\end{equation}

Bonnor \cite{Bonnor} proposed to eliminate the ambiguities 
that remain defining the passive gravitational mass as:

\begin{equation}
\label {3.3}
M(B)=4\pi\int_0^\infty{(\rho+p){\tilde B}{\tilde r}^2}\, d\tilde r
\end{equation}
This means that $\rho+p$ is interpreted as a density of
inertial, or passive, gravitational mass and that the metric that gives a meaning to the
word volume is (\ref{1.2}). But this metric is known only once the problem has been
solved and then it depends on the point of the body which is considered.
This deprives the meaning of the variables $\rho$ and $p$ and the
compressibility equation (\ref{1.16}) of any {\it a
priori} significance. Other difficulties that arise from this
definition are discussed in Bonnor's paper.

We do not believe that the consideration of a single body at rest, as we
have been doing here, can say anything about its passive or
inertial
gravitational mass because this would require to know how it reacts to the
presence of another comparable body. On the other hand we believe that we
should be able to define its proper mass $M_p$ if we want to know what is the
binding energy $E_b$ of any given configuration as defined by:

\begin{equation}
\label {3.3.1}
E_b=M_a-M_p
\end{equation}

More precisely, our point of view, along the lines of a long enduring effort to
understand the concept of rigidity and establishing a theory of
frames of reference in special and general relativity, consists in
accepting the usual generalization of Schwarzschild's ``substantial mass" \, 
\footnote{An english translation of Schwarzschild's paper has been provided 
by S. Antoci in arXiv:physics/9912033} as proper mass:

\begin{equation}
\label {3.4}
M_p=4\pi\int_0^\infty{\rho B^2Cr^2}\, dr, \quad \sqrt{\hat g}=B^2Cr^2\sin\theta
\end{equation}
and defining at the same time a proper mass density $\rho_p$:

\begin{equation}
\label {3.3.2}
\rho_p = \rho B^2C  
\end{equation}
such that $M_p$ could be written:

\begin{equation}
\label {3.6}
M_p=4\pi \int^\infty_0 {\rho_p r^2}\,dr, \quad \sqrt{\bar g}=r^2\sin\theta
\end{equation}
This means then defining $\rho_p$ as a density of proper
gravitational mass and interpreting the words volume and
surface in the sense of the universal euclidian geometry (\ref{1.10})
related to the quotient metric (\ref{1.2}) by a principal transformation
(\ref{1.6}). This guarantees the independence of the meanings of $\rho_p$,
$p$ and the compressibility equation (\ref{1.16}) independently of the
location of the element of the fluid in the object and independently of
the solution of the field equations that one is considering. This
guarantees also that $M_p$ can be identified with an appropriate number of identical
samples of a fluid as weighted with a balance at the ``shop store"
before being assembled into the body.

The interpretation we have just given of the metric (\ref{1.10}) implies
also as a corollary that, as above-mentioned, a fluid with isotropic
pressure should be described by an energy-momentum tensor as written
in (\ref{1.13}). And that for the compressibility equation to have a
well defined {\it a priori} meaning it should be given as a
relationship between $\rho_p$ and $p$:

\begin{equation}
\label {3.4.1}
\rho_p=\rho_p(p) \quad \hbox{or} \quad p=p(\rho_p)
\end{equation}

\section{Localized energy density}

Following suit to the ideas of the preceding section we exhibit the
quantity $E_b$ as an integral extended over all
space of an energy density function depending only on $r,B,C$ and
$B^\prime,C^\prime$. From (\ref{3.2}) and (\ref{3.6}) it follows
that:

\begin{equation}
\label {4.1}
E_b=4\pi\int_0^\infty{\rho B^3Yr^2}\,dr
\end{equation}
where:

\begin{equation}
\label {4.1bis}
Y=\frac{F}{B^3}-\frac{C}{B}
\end{equation}
while $\rho B^3$ can be obtained from (\ref{2.1}) as:

\begin{equation}
\label {4.2}
\rho B^3=\frac{1}{8\pi}(X-B^{\prime\prime})
\end{equation}
with:

\begin{equation}
\label {4.3}
X=\frac{B(-3rB^\prime-B+B^2/C)}{r^2C}+\frac54\frac{{B^\prime}^2}{B}
\end{equation}
Therefore we have:

\begin{equation}
\label {4.4}
E_b=\frac12\int_0^\infty{XYr^2}\,dr
-\frac12\int_0^\infty{B^{\prime\prime} Yr^2}\,dr
\end{equation}
Integrating by parts the second integral we obtain:

\begin{equation}
\label {4.5}
E_b=4\pi\int_0^\infty{\sigma(r)r^2}\,dr-
4\pi\lim_{r\to\infty}\sqrt{\frac{C}{B}}B^\prime r^2
\left(\frac14\frac{B^\prime Cr}{B^2}+1-\sqrt{\frac{C}{B}}\right)
\end{equation}
where:

\begin{equation}
\label {4.6}
\sigma(r)=\frac{1}{8\pi}(XY+Z)
\end{equation}
with:

\begin{equation}
\label {4.6.1}
Z=\frac{B^\prime}{8rB^3\sqrt{BC}}(-5r^2C^2{B^\prime}^2
+2rBC(4\sqrt{BC}+C){B^\prime}-8B^3(2\sqrt{BC}-B+C))
\end{equation}
depends only on $r,B,C$ and $B^\prime$.
From the asymptotic conditions (\ref{1.17}) it follows that :

\begin{equation}
\label {4.7}
\lim_{r\to\infty}B/C=1 \quad \hbox{and} \quad  \lim_{r\to\infty}r^3{B^\prime}^2=0
\end{equation}
and therefore, the limit in (\ref{4.5}) being zero, the final result is:

\begin{equation}
\label {4.8}
E_b=4\pi\int_0^\infty{\sigma(r)r^2}\,dr
\end{equation}

\section{Linear approximation}
We consider here the linear approximation  of the models that we have been
considering, to take a closer look to two of the de concepts that
we have implemented in Sect. 3. Namely: the mass defect, or binding energy $E_b$ and the
proper mass density $\rho_p$.

We assume that $A$, $B$ and $C$ can be written as:

\begin{equation}
\label {5.1}
A=1+A_1, \quad B=1+B_1, \quad C=1+C_1
\end{equation}
where $A_1$, $B_1$ and $C_1$ are small quantities, of order
$\epsilon$ say. We assume also that $\rho$ is also of order $\epsilon$
and that $p$ is of order $\epsilon^2$ and can be ignored, as well as any other 
quantity of the same order or
smaller, in the field equations. 

The coordinate condition (\ref{1.5}) and the field equations
(\ref{2.1}), (\ref{2.2}) 
become then:

\begin{eqnarray}
\label {5.2}
C_1^\prime&=&\frac{2}{r}(B_1-C_1)\\
B_1^{\prime\prime}&=&-3\frac{B_1^\prime}{r}+\frac{1}{r^2}(B_1-C_1)-8\pi
\rho\\
A_1^{\prime\prime}&=&\frac{1}{2r}(B_1^\prime-2A_1^\prime)
+\frac{1}{2r^2}(B_1-C_1)+4\pi\rho
\end{eqnarray}

As our purpose is purely illustrative here  we consider below the
simplest case where the source is a spherical body of finite radius
$R$ and constant $\rho$. The interior solution satisfying the
regularity conditions (\ref{1.18}) at the centre is:

\begin{equation}
\label {5.3}
A_1=\frac{2}{3}\pi\rho r^2 + a_0, \quad B_1=-\frac{16}{15}\pi\rho r^2+ b_0
\quad C_1=-\frac{8}{15}\pi\rho r^2+b_0
\end{equation}
where $a_0$ and $b_0$ are two allowed constants of integration;
the exterior solution satisfying the asymptotic conditions
(\ref{1.17}) at infinity is:

\begin{equation}
\label {5.4}
A_1=-\frac{b_1}{r}, \quad B_1=\frac{b_1}{r}+\frac{b_3}{r^3}, \quad
C_1=\frac{2b_1}{r}-\frac{2b_3}{r^3}
\end{equation}
where $a_1$ and $b_3$ are two new constants of integration. Demanding
the continuity of $A_1$, $B_1$ and $C_1$ and their first derivatives
across the border $r=R$ fixes $a_0$ and all the $b$'s as follows:

\begin{equation}
\label {5.5}
a_0=-2\pi\rho R^2, \quad b_0=\frac83\pi\rho R^2, \quad
b_1=\frac43\pi\rho R^3, \quad b_3=\frac{4}{15}\rho R^5
\end{equation} 

From the preceding results we can calculate the leading approximation, 
which is of order $\epsilon^2$, of the localized energy density
(\ref{4.6}).
For $r<R$ the result is using an arbitrary system of units:

\begin{equation}
\label {5.6}
\sigma=-\frac{3}{20}\frac{G^2M^2r^2}{\pi R^6}
\end{equation} 
where at this approximation $M$ is either $M_a$ or $M_p$. For $r>R
$ the result is:

\begin{equation}
\label {5.7}
\sigma=-\frac{3}{160}\frac{G^2M^2(5r^4+6R^2r^2-3R^4)}{\pi r^8}
\end{equation}

The binding energy can be calculated using (\ref{4.8}), (\ref{5.6}) and
(\ref{5.7}), 
or (\ref{3.3.1}), (\ref{3.2}) and (\ref{3.6}) at the appropriate approximation. 
The result is, using arbitrary units, the familiar Newtonian amount:

\begin{equation}
\label {5.7.1}
E_b=-\frac{3}{5}\frac{GM^2}{c^2R},
\end{equation}
a result that can be obtained using a variety of other approaches\,
\footnote{See for instance reference \cite{HerrIba}}
\section{Massive Neutron Cores}

Any particular model will be characterized by an equation of state and
the value of its central density, or central pressure, or both in the
important case in which one assumes that the density is constant.
Taking into account the regularity conditions 
the initial conditions of the gravitational potentials $A, B$ and $C$ 
have to be chosen such that:

\begin{equation}
\label {6.1}
B_0=C_0, \quad A^\prime_0=B^\prime_0=C^\prime_0.
\end{equation}
And taking into account the asymptotic conditions the values of $A_0$ and 
$B_0$ have to be chosen such that:

\begin{equation}
\label {6.2}
\lim_{r\to\infty}A=\lim_{r\to\infty}B=1
\end{equation}  
the condition:

\begin{equation}
\label {6.3}
\lim_{r\to\infty}C=1
\end{equation}
as, well as the remaining asymptotic conditions, being then automatically satisfied 
because the solution behaves as the exterior Schwarzschild one at infinity. 

The numerical integration of the system of equations (\ref{1.5}),
(\ref{2.1}), (\ref{2.2}) and (\ref{2.5}) where:

\begin{equation}
\label {6.4}
\rho=\rho_p(f(B^2C)^{-1}+1-f)
\end{equation}
and $\rho_p$ is a known function of $p$, is a trial and error
procedure. Arbitrary values of $A_0, B_0$ and $p_0>0$ have to be
chosen; the integration has to proceed until $p=0$; then the equation
of state has to be abandoned and $\rho=p=0$ has to be required; the
integration has then to proceed to sufficiently large values of $r$ to
check the asymptotic conditions (\ref{6.2}). If the check is not
satisfactory the whole process has to be started again with new
values of $A_0, B_0$ and $p_0>0$.

As an important example we consider the equation of state of a
degenerate neutron gas as it suits to a model of massive neutron
cores. This was considered in a famous paper by Oppenheimer and
Volkoff from the standard point of view which consists in putting
$f=0$. Our point of view consist in using instead the value $f=1$.

The equation of state can be written in parametric form, including 
both points of view, as:

\begin{eqnarray}
\label {6.5}
\rho&=&K(\sinh u -u)(f(B^2C)^{-1}+1-f) \\
p&=&\frac13 K(\sinh u-8\sinh \frac12 u+3u)
\end{eqnarray}     
where,using arbitrary units:

\begin{equation}
\label {6.6}
K=\frac{m^4c^5}{4h^3}
\end{equation} 
$m$ being the mass of a neutron.

We recall below the values of $M_a$ obtained in \cite{OppenVolkoff}
for several values of the inial values of $u_0$, and include the values of 
$M_p$ as calculated from (\ref{3.4}):

\vspace{0.5cm}

\begin{tabular}{|l|l|l|l|}
\hline
$u_0$ & $M_a$ & $M_p$ & $E_b$ \\
\hline
1 & 0.033 & 0.033 & -0.000 \\
2 & 0.066 & 0.071 & -0.005 \\
3 & 0.077 & 0.088 & -0.011 \\
4 & 0.070 & 0.085 & -0.015 \\
5 & 0.060 & 0.074 & -0.014 \\
6 & 0.049 & 0.062 & -0.013 \\
7 & 0.042 & 0.054 & -0.012 \\
\hline
\end{tabular}

\vspace{0.5cm}
The system of units that has been used is that satisfying
(\ref{1.19})
completed with the supplementary condition:

\begin{equation}
\label {6.7}
\bar K=1/4\pi
\end{equation}
which is the choice made in \cite{OppenVolkoff} to define
a unit of mass.  The most notorious result is existence of a maximum
mass corresponding approximately to $u_0=3$ whose value is $M_a=0.078$
which corresponds to $M_a=0.71\,M_\odot $. Oppenheimer and Volkoff
concluded also from a very crude non relativistic argument that above
$u_0=3$ the equilibrium configurations were not stable. 

For $f=1$ the results that we have obtained are:
\vspace{0.5cm}

\begin{tabular}{|l|l|l|l|}
\hline
$u_0$ & $M_a$ & $M_p$ & $E_b$ \\
\hline
1 & 0.04 & 0.04 & -0.00 \\
2 & 0.12 & 0.13 & -0.01 \\
3 & 0.24 & 0.28 & -0.04 \\
4 & 0.36 & 0.47 & -0.11 \\
5 & 0.41 & 0.56 & -0.15 \\
6 & 0.36 & 0.51 & -0.15 \\
7 & 0.33 & 0.48 & -0.15 \\
\hline
\end{tabular}

\vspace{0.5cm}

Here also we obtain that there is a maximum mass $M_a=0.41$ which
corresponds to $M_a=3.8\,M_\odot$ but it is substantially larger than
the value in \cite{OppenVolkoff} as well as the mass of some models 
with anisotropic pressure considered by Corchero 
in \cite{Corchero} . It is even somewhat larger than the limit
value, $M_a=3.2\,M_\odot$ obtained by Rhoades and Ruffini in
\cite{RhoaRuff} from very general considerations complying with the
conventional point of view. 

\section*{Acknowledgments}

I gratefully acknowledge the help provided by J.~M.~Aguirregabiria
checking parts of this manuscript, the patience of A.~Chamorro for a careful reading
of it, and their comments as well as those of E.~S.~Corchero and J.~Mart\'{\i}n. 
I also gratefully acknowledge the position of
visiting professor to the UPV/EHU that I have been holding while this
paper was being prepared.



 
\end{document}